\documentstyle[prc,aps]{revtex}
\newfont{\boldit}{cmbxti10 at 14 pt}

\begin{document}
\title{Realistic shell-model calculations for proton-rich {\boldit
N}{\bf=50} isotones}
\author{L. Coraggio$^{1}$, A. Covello$^1$,
A. Gargano$^1$, N. Itaco$^1$, and T. T. S. Kuo$^{2}$}
\address{$^1$Dipartimento di Scienze Fisiche, Universit\`a
di Napoli Federico II, \\ and Istituto Nazionale di Fisica Nucleare, \\
Complesso Universitario di Monte S. Angelo, Via Cintia, I-80126 Napoli,
Italy \\
$^2$Department of Physics, SUNY, Stony Brook, New York 11794}
\date{\today}

\maketitle

\begin{abstract}
The structure of the $N=50$ isotones $^{98}$Cd, $^{97}$Ag, and $^{96}$Pd is
studied in terms of shell model employing a realistic effective interaction
derived from the Bonn-A nucleon-nucleon potential. The single-hole energies
are fixed by resorting to an analysis of the low-energy spectra of the
isotones with $A \geq 89$. Comparison shows that our results are in very
satisfactory agreement with the available experimental data. This supports
confidence in the predictions of our calculations.

\end{abstract}

\draft

\section{Introduction}

In recent years, we have studied a number of nuclei around doubly magic
$^{100}$Sn, $^{132}$Sn, and $^{208}$Pb
\cite{andr96,andr97,cov97,cov98,cor98,andr99,cov99,cor99}
within the framework of the shell model employing realistic effective
interactions derived from the meson-theoretic Bonn-A nucleon-nucleon ($NN$)
potential \cite{machl87,machl89}. We have focused attention on nuclei with
few valence particles or holes, since they provide the best testing ground
for the basic ingredients of shell-model calculations, especially as regards
the matrix elements of the two-body effective interaction. The main
motivation for carrying out this extensive program of calculations was to
try to assess the role of realistic effective interactions in the
shell-model approach to the description of nuclear structure properties. The
results of our calculations have so far turned out to be in remarkably good
agreement with experiment for all the nuclei considered, providing evidence
that realistic effective interactions are able to describe with quantitative
accuracy the spectroscopic properties of complex nuclei. In this connection,
it is worth noting that these results are considerably better than those
obtained in earlier works for the light $s$-$d$ nuclei
\cite{jiang92,hjorth95}.

While in the $^{132}$Sn and $^{208}$Pb regions we have studied nuclei with
both valence particles and holes, around $A=100$ we have only considered
the light Sn isotopes, namely we have not gone below the 50-82 shell.
The study of nuclei lacking few nucleons with respect to $^{100}$Sn, which
is the heaviest $N=Z$ doubly magic nucleus, is of course of great relevance
from the shell-model point of view. Nuclei of this kind, however, lie well
away from the valley of stability and experimental information on their
spectroscopic properties is still very scanty. In this context, the
proton-rich $N=50$ isotones are of special interest. In fact, while the
development of radioactive ion beams opens up the prospect of spectroscopic
studies of a number of $^{100}$Sn neighbors, use of large multidetector
$\gamma$-ray arrays is already providing new experimental data for these
singly magic nuclei. In particular, four excited states in $^{98}$Cd, two
proton holes from $^{100}$Sn, have been recently identified in an in-beam
spectroscopy experiment \cite{gorska97}.

On the above grounds, we found it very interesting to include in our program
of calculations the proton-rich $N=50$ isotones $^{98}$Cd, $^{97}$Ag, and
$^{96}$Pd (some preliminary results have already been presented in Ref.
\cite{cov98}). Actually, the motivation for the present study is twofold. On
the one hand, these nuclei with two, three, and four holes in the $Z=28-50$
shell offer the opportunity to test our realistic effective interaction for
nuclei below $^{100}$Sn. On the other hand, the success achieved by our
previous calculations on medium- and heavy-mass nuclei encourages us to make
predictions which may stimulate, and be helpful to, future experiments.

The $N=50$ isotones have long been the subject of theoretical interest.
In most of the shell-model calculations performed in the last two decades
\cite{blomq85,ji88a,ji88b,sinatk92,ghugre95,johnst97,zhang99}
(earlier references can be found in Ref. \cite{blomq85}), however, attention
has been focused on the lighter isotones up to mass 95
\cite{ghugre95} or 96\cite{ji88a,ji88b,sinatk92,johnst97,zhang99}. In this context, we may mention the extensive study of the $N=50$ isotones from $^{82}$Ge up to $^{96}$Pd performed some ten years ago by Ji and Wildenthal \cite{ji88a,ji88b}. In that work $^{78}$Ni was considered as a closed core and an empirical effective Hamiltonian was obtained by fitting the two-body matrix elements and the single-particle energies to approximately 170 experimental data. 
Actually, the low-energy spectra of $^{98}$Cd and $^{97}$Ag have been
predicted only in the work of Ref. \cite{blomq85}, where the protons were
assumed to fill solely the $0g_{9/2}$ and $1p_{1/2}$ levels.

In all previous calculations empirical two-body matrix elements have been used, an exception being the work of  Ref. \cite{sinatk92}, where the effective interaction was derived from the Sussex interaction \cite{elliott68}. To our knowledge, the present calculations are the first ones where the two-body effective interaction has been derived from a modern free
nucleon-nucleon potential. 

Before closing this section we should remark that, at variance with our
previous calculations in the $^{132}$Sn and $^{208}$Pb regions, we
had to face here the problem of choosing a set of single proton-hole
energies without much guidance from experiment. In fact, while no
spectroscopic data are yet available for the
single-hole valence nucleus $^{99}$In, only little relevant information is
provided by the observed spectra of $^{98}$Cd and $^{97}$Ag.
We will come back to this important point later.

The paper is organized as follows. In Sec. II we give an outline of our
calculations and describe in detail how we have determined the single-hole
energies. Our results are presented and compared with the experimental data in Sec. III,
where we also comment on the results of Ref. \cite{sinatk92}. Section IV presents a summary of our conclusions.

\section{Outline of calculations}

Our  effective interaction $V_{\mathrm eff}$ was derived
from the Bonn-A free nucleon-nucleon ($NN$) potential using
a $G$-matrix formalism, including
renormalizations from both core polarization and folded diagrams.
Since we have assumed $^{100}$Sn as a closed core,
protons are treated as valence holes, which implies the derivation of a
hole-hole effective interaction.
We have chosen the Pauli exclusion operator $Q_2$ in the $G$-matrix
equation,

\begin{equation}
G(\omega)=V+V Q_2 {{1} \over{\omega-Q_2TQ_2}} Q_2G(\omega),
\end{equation}

\noindent
as specified \cite{krenc76,kuo96} by ($n_1, n_2, n_3$) = (11, 21, 55) for
both
neutron and proton orbits.
Here $V$ represents the $NN$ potential, $T$ denotes the two-nucleon kinetic
energy, and $\omega$ is the so-called starting energy.
We employ a matrix inversion method to calculate the above $G$ matrix in an
essentially exact way \cite{tsai72}.
For the harmonic oscillator parameter $\hbar \omega$ we adopt the value
8.5 MeV, as given by the expression $\hbar \omega=45A^{-1/3}-25A^{-2/3}$ for
$A=100$.
Using the above $G$ matrix we then calculate the so-called
$\hat Q$ box, which is composed of irreducible valence-linked
diagrams up to second order in $G$. These are just the seven
one- and two-body diagrams considered
in Ref. \cite{shurp83}.

Since we are dealing with external hole lines, the calculation of the
$\hat{Q}$-box diagrams is somewhat different from that usual for
particles.
For example, the familiar core-polarization diagram $G_{\mathrm 3p1h}$
becomes

\begin{eqnarray}
\langle a~b;J| G_{\mathrm 3h1p} |c~d;J \rangle & = &
-\frac{(-1)^{j_a+j_b+j_c+j_d}} {\hat{J}} \sum_{J'} \hat{J'} \sum_{\mathrm
ph}
(-1)^{j_p-j_h+J'}
X \left( \begin{array}{ccc}
j_c~ j_d~  J  \\ 
j_a~ j_b~  J  \\ J'~  J'~  0  \end{array} \right) \nonumber \\
&& \mbox{} \times \frac{1}{\omega
-(\epsilon_p-\epsilon_h-\epsilon_b-\epsilon_c)}
 \langle~ \rule[-0.15in]{0.65in}{0.01in} \hspace{-0.68in} \raisebox
{-0.13in}
{$ \scriptstyle \uparrow$} \hspace{-0.05in}c ~
\rule [0.21in]{0.65in}{0.01in}\hspace{-0.68in}
\raisebox {0.14in} {$ \scriptstyle \downarrow$} \hspace{-0.05in} h
| G(\omega_{1}) \raisebox{0.25in}{$\scriptstyle J'$}
\hspace{-0.15in}
\raisebox{-0.25in}{$\scriptstyle J'$}
\hspace{-0.05in}|~
 \raisebox{-0.12in}{$\scriptstyle \mid$} \hspace{-0.05in} a \hspace{0.01in}
 \raisebox{0.14in}{$\scriptstyle \mid$} \hspace{-0.04in}p \rangle
\langle~ \rule[0.21in]{0.67in}{0.01in}\hspace{-0.70in} \raisebox {0.14in}
{$\scriptstyle \downarrow$} \hspace{-0.05in}p ~
\rule[-0.15in]{0.67in}{0.01in}\hspace{-0.70in} \raisebox{-0.13in}
{$\scriptstyle \uparrow$} \hspace{-0.05in} d
| G(\omega_{2}) \raisebox{0.25in}{$\scriptstyle J'$}
\hspace{-0.15in}
\raisebox{-0.25in} {$\scriptstyle J'$}
\hspace{-0.05in}|~
\raisebox{0.14in}{$\scriptstyle \mid$} \hspace{-0.06in}h~ \hspace{0.00in}
 \raisebox{-0.12in}{$\scriptstyle \mid$} \hspace{-0.06in}
 b \rangle ,
\end{eqnarray}

\noindent
where $\hat{x}=(2x+1)^{1/2}$ and the off-shell energy variables are:
$\omega_1=\omega+\epsilon_h+\epsilon_a+\epsilon_b+\epsilon_c$ and
$\omega_2 =\omega+\epsilon_h+\epsilon_b+\epsilon_c+\epsilon_d$.
The $\epsilon$'s are the unperturbed single-particle energies.
$X$ is the standard normalized 9-$j$ symbol.
The cross-coupled $G$-matrix elements, on the right side of Eq. (2), are
related to the usual direct-coupled ones by a simple transformation, as in
Ref.\cite{kuo81}.

The effective interaction, which is energy independent, can be schematically
written in operator form as:

\begin{equation}
V_{\mathrm eff}= \hat{Q} - \hat{Q}^{'}\int\hat{Q}
+ \hat{Q}^{'}\int\hat{Q}\int\hat{Q} - \hat{Q}^{'}\int\hat{Q}\int\hat{Q}
\int\hat{Q}\cdots,
\end{equation}

\noindent
where the integral sign represents a generalized folding operation
\cite{krenc80}.
$\hat{Q}^{'}$ is obtained from $\hat{Q}$ by removing terms of first order in
the reaction matrix $G$.
After the $\hat{Q}$ box is calculated, $V_{\mathrm eff}$ is then obtained by
summing up the folded-diagram series of
Eq. (3) to all orders using the Lee-Suzuki iteration method \cite{lesu80}.

As regards the electromagnetic observables, these have been calculated
by making use of effective operators \cite{mavro66,krenc75} which
take into account core-polarization effects.
More precisely, by using a diagrammatic description as in Ref.
\cite{mavro66},
we have only included first-order diagrams in $G$.
This implies that folded-diagram renormalizations are not needed
\cite{krenc75}.

Let us now come to the single-hole (SH) energies. As already mentioned in
the Introduction, no spectroscopic information on $^{99}$In is available.
To obtain information on  the location  of the SH
levels we have therefore  resorted to an analysis of the spectra
of the lighter $N=50$ isotones. Of course, most relevant to this analysis
are those states which are predominantly of one-hole nature.
Actually, $^{91}$Nb is the first isotone where a state of this kind has been
unambiguously identified for each of the four SH levels.
More precisely, no states with a firm
$\frac{3}{2}^{-}$ or $\frac{5}{2}^{-}$ assignment are reported for
the heavier isotones, while at least one $\frac{1}{2}^{-}$  and one
$\frac{9}{2}^{+}$ state have been identified up to $^{95}$Rh. In $^{97}$Ag
there is only one $\frac{9}{2}^{+}$ state, which is the ground state.

>From the above it is clear that, if one wants to
determine the SH energies by reproducing the observed one-hole
states, calculations up to
nine valence holes have to be carried out.
It is to be expected, however, that a set of  SH energies determined in
this way may not be
the most appropriate for calculations where few valence holes are
considered.
In fact, as is well known, significant changes in the nuclear mean field
may occur when moving away  from closed shells. In addition,
an effective two-hole interaction derived by considering $^{100}$Sn as a
closed core may not be quite adequate for systems with several valence holes
as, in these cases, many-body correlations are likely to come into play.

In this situation,  we have tried  to determine the SH energies
$\epsilon_{p_{1/2}}$, $\epsilon_{p_{3/2}}$, and  $\epsilon_{f_{5/2}}$
relative to the
$g_{9/2}$ level, which has long been recognized to be the lowest-lying one,
from an analysis of  the spectra of the $^{100}$Sn neighbors $^{98}$Cd,
$^{97}$Ag, and $^{96}$Pd, with two to four proton holes in the $N=28-50$
shell.
We have found that
(i) the energies
of all the excited levels in $^{98}$Cd and $^{97}$Ag, which have an
experimental counterpart, are quite insensitive to the
position of the $p_{3/2}$ and $f_{5/2}$ orbits; (ii) the ground-state
energies of all three nuclei, as well as the seniority-two states
$J^{\pi}=2^{+},4^{+},6^{+},$ and $8^{+}$ in $^{96}$Pd,
depend practically only on
the sum of the energies of these two levels,
$\epsilon=\epsilon_{p_{3/2}}+\epsilon_{f_{5/2}}$.
It turns out that all the considered experimental spectra  are
well described overall by fixing $\epsilon$ at 5.2 MeV.
More precisely, only the $2^+$ states in the two even isotones and the
$\frac{13}{2}^+$ and the $10^+$ states in $^{97}$Ag and $^{96}$Pd show
a rather large discrepancy. To eliminate this discrepancy a much larger
value of $\epsilon$ should be used, namely about 10 MeV.
This value, however, would
produce a significant downshift of all other levels. In addition,
as we shall see in the following, it
would be at variance with an empirical
analysis of the one-hole states in $N=50$ isotones. It may also be
mentioned that the energies  of the  $2^+$
states, as well as those of the $\frac{13}{2}^+$ and $10^+$ states,
are all strongly dependent on the two-body matrix element
$\langle g_{9/2}^{-2} J^{\pi}=2^{+}|V_{\rm eff}|g_{9/2}^{-2} J^{\pi}=2^{+}
\rangle$. In this context, we  should recall  that
also for the light
Sn isotopes our calculations with the Bonn-A potential produced
$2^+$ excitation energies somewhat higher than the observed
values \cite{andr96}.

As for the $p_{1/2}$ level, two states are sensitive to its
position. They are the $\frac{17}{2}^{-}$ and $5^-$ states in $^{97}$Ag and
$^{96}$Pd, respectively. We find that their experimental  energies
are very well reproduced by our calculations for
$\epsilon_{p_{1/2}}=0.7$ MeV. We have verified that this choice is
rather independent of the
value of $\epsilon$. For instance,  increasing
$\epsilon$ by about 2 MeV brings $\epsilon_{p_{1/2}}$ up to only
0.9 MeV.

>From the above findings it appears that the SH energies
$\epsilon_{p_{3/2}}$ and $\epsilon_{f_{5/2}}$
cannot be determined individually from the
experimental data for $^{98}$Cd, $^{97}$Ag, and $^{96}$Pd presently
available.
To obtain an estimate for these two $\epsilon$'s,
we have made a linear extrapolation  of the
energies of the $\frac{3}{2}^-$ and $\frac{5}{2}^-$ one-hole states
observed in $^{89}$Y, $^{91}$Nb, and  $^{93}$Tc. Actually,
states of this kind have been unambiguously identified only
in $^{89}$Y and $^{91}$Nb. In particular , in the latter nucleus two
$\frac{3}{2}^{-}$ states have been observed which exhaust almost all the
$p_{3/2}$ strength. In our extrapolation, however, 
we have also included the experimental data relative 
to $^{93}$Tc, according to the indications of Ref. \cite{shamai72}. In
this work the level at
2.1 MeV is identified as an $l=3$, $J=\frac{5}{2}$ state while   
plausible arguments are given favoring the  $\frac{3}{2}^{-}$ 
assignment to  the
two states observed at 1.5 and 1.8 MeV.
The above procedure yields the
values of about 2 and 3 MeV for the  $\frac{3}{2}^{-}$ and
$\frac{5}{2}^{-}$ SH energies in $^{99}$In.
Owing to the uncertainty inherent in
such a derivation, these values should be taken only
as a reasonable estimate. In support of this procedure, however, speaks the
fact that for
the $p_{1/2}$ level it yields $\epsilon_{p_{1/2}}=0.8$ MeV.

On the above grounds, we have adopted for the SH energies the following
values (in MeV):
$\epsilon_{g_{9/2}}=0.0$, $\epsilon_{p_{1/2}}=0.7$,
$\epsilon_{p_{3/2}}=2.1$,
and $\epsilon_{f_{5/2}}=3.1$.
It should be pointed out that these values are quite different
from those adopted by other authors. In particular, the SH energies
determined
in Ref. \cite{sinatk92} are higher than  ours, the difference  ranging from
more than 1 MeV for $\epsilon_{p_{1/2}}$ and $\epsilon_{p_{3/2}}$  to 3.2
MeV
for $\epsilon_{f_{5/2}}$.

\section{Results and Comparison with experiment}

We present here the results of our calculations for $^{98}$Cd, $^{97}$Ag, and
$^{96}$Pd. They have been obtained by using the OXBASH shell-model code
\cite{brown}.  
The experimental \cite{nndc} and theoretical spectra 
are compared in Figs. 1, 2, and 3, where we report all the 
experimental levels,
except the $13^+$ and $15^+$
states observed at 6.7 and 7.0 MeV in $^{96}$Pd,
which cannot be constructed in our model space. In the calculated
spectra only those yrast states which are candidates for the observed 
levels
are reported. A complete list of excitation energies
up to 5, 3, and 4 MeV is given in Tables I-III for $^{98}$Cd $^{97}$Ag,
and $^{96}$Pd, respectively.

\renewcommand{\thefootnote}{\fnsymbol{footnote}}
\setcounter{footnote}{1}

>From Figs. 1-3 we see that our results are in very good agreement with
experiment.
A measure of the quality of the agreement is given by the rms deviation
$\sigma$\footnote{We define
$\sigma =\{(1/N_{d}) \sum_{i}[E_{exp}(i)-E_{calc}(i)]^{2}\}^{1/2}$, where
$N_{d}$ is the number of data.}, 
whose values
are 107, 108, and 122 keV for $^{98}$Cd, $^{97}$Ag, and
$^{96}$Pd, respectively. As was already discussed in
Sec. II,
the main point of disagreement is the position of the
$2^+$ state in both the even isotones, as well as that of the
$\frac{13}{2}^+$
and $10^+$ states in $^{97}$Ag and $^{96}$Pd. In fact, the discrepancy
between theory and experiment for the energies of these four states
goes from 140 to 263 keV while it is less than 100 keV for all
other states.

As regards the structure of the states having an experimental 
counterpart,
we find that the
positive-parity states in all three nuclei are dominated
by the $g_{9/2}^{-n}$ configuration, while the negative parity ones are
practically of pure $g_{9/2}^{-(n-1)}p_{1/2}^{-1}$ character. In
$^{98}$Cd and $^{97}$Ag only the ground states receive a
significant contribution from configurations other than the
dominant one, the percentage being about 20\%
in both nuclei. As for $^{96}$Pd, the wave functions of the
ground state and the first four excited states
are even less pure. In fact,  the percentage
of the $g_{9/2}^{-4}$ configuration reaches at most
81\% for the $4^{+}_{1}$, $6^{+}_{1}$, and $8^{+}_{1}$ states, being
only 64\% for the ground state.
Note that in these states, as well as in the ground states of
$^{98}$Cd and $^{97}$Ag, a significant percentage of the
$g_{9/2}^{-(n-2)}p_{3/2}^{-2}$ and $g_{9/2}^{-(n-2)}f_{5/2}^{-2}$ is
present.
In particular, in the ground state of $^{96}$Pd the percentage of each of
these two these configurations is 9\%.

>From Figs. 1-3 we see
that rather little experimental information is presently available for
$^{98}$Cd, $^{97}$Ag, and $^{96}$Pd. Much richer spectra, however,
are predicted by the theory. It is therefore interesting
to discuss in some detail our predictions, in the hope that they may
verified in a not too distant future.
As regards $^{98}$Cd, it may be seen from
Table I that, just above the first four excited
states having an experimental counterpart, we find 
three states with $J^{\pi}=4^{-}, 5^{-}$, and $0^+$, the first two
being the  members of the
doublet $g_{9/2}^{-1}p_{1/2}^{-1}$ and the third one arising 
from the configuration $p_{1/2}^{-2}$.
The position
of the $5^-$ state is quite consistent with the experimental information
available for the two lighter even isotones. In fact, in $^{96}$Pd and
$^{94}$Ru a $5^-$ state has been observed at 2.65 and 2.62 MeV,
respectively.
Between 3.8
and 5 MeV we find all the members of the $g_{9/2}^{-1}p_{3/2}^{-1}$
multiplet
and the $7^-$ state arising from the $g_{9/2}^{-1}f_{5/2}^{-1}$
configuration. In this energy interval is also located the $2^+$
state of the
$p_{1/2}^{-1}p_{3/2}^{-1}$ configuration. 

In Table II all the excitation energies up to 3 MeV are reported for
$^{97}$Ag. Below this energy we find all the states arising 
from the configurations
$g_{9/2}^{-3}$, $g_{9/2}^{-2}p_{1/2}^{-1}$, and $g_{9/2}^{-1}p_{1/2}^{-2}$
as well as the two seniority-one states of the $g_{9/2}^{-2}p_{3/2}^{-1}$ 
and $g_{9/2}^{-2}f_{5/2}^{-1}$ configurations.
In particular, we predict as first excited state a seniority-one
$\frac{1}{2}^-$ state at about 0.5 MeV. This prediction is in agreement
with the experimental findings for the lighter isotones \cite{nndc}.
Furthermore, it should be mentioned that our
first $\frac{5}{2}^-$ state is essentially a pure seniority-three
$g_{9/2}^{-2}p_{1/2}^{-1}$ state while almost all the $l=3$ one-hole
strength is concentrated in the second one at 2.6 MeV. On the other hand,
we find that the $p_{3/2}$ strength is almost equally distributed between
the first and second $\frac{3}{2}^-$ states.

As for $^{96}$Pd, only 2 out of the 25 states which we predict
up to 4 MeV (see Table III) arise from
configurations other than $g_{9/2}^{-4}$ and $g_{9/2}^{-3}p_{1/2}^{-1}$.
They are the $J^{\pi}=0^{+}_{2}$ and
$3^{-}_{1}$ states with a $g_{9/2}^{-2}p_{1/2}^{-2}$ and
$g_{9/2}^{-3}p_{3/2}^{-1}$ dominant component, respectively.

>From the above discussion it is evident that some of our predictions are closely related to the values adopted for $\epsilon_{p_{3/2}}$ and $\epsilon_{f_{5/2}}$.
For instance, as shown before, we find that the wave functions of several states in the three considered isotones contain non negligible components outside the ($g_{9/2}$,$p_{1/2}$) space. This indicates that a two-level model space would not be adequate even for the description of the heavier
$N=50$ isotones. We also predict the absence of a pronounced gap above the
$0^{+}_{2}$ state in the spectrum of $^{98}$Cd as well as rather low-lying 
one-hole $\frac{3}{2}^{-}$ and $\frac{5}{2}^{-}$ states in $^{97}$Ag. 
This makes it clear that, in absence of a spectroscopic study of $^{99}$In, 
the discovery of new selected levels in $^{98}$Cd and $^{97}$Ag represents the best source of information on the SH spectrum.

To conclude this discussion, a further comment is in order.
As it occurs for the $\frac{13}{2}^+$ state in $^{97}$Ag and the $2^+$ and
$10^+$ states in $^{96}$Pd, we expect that the calculated excitation
energies of all other states in these two nuclei
arising from the $2^+$ state of $^{98}$Cd may be
somewhat overestimated (200-300 keV). This is the case,
for instance, of the
$(\frac{5}{2}^{+})_{1}$, $(\frac{7}{2}^{+})_{1}$, $(\frac{9}{2}^{+})_{2}$,
and $(\frac{11}{2}^{+})_{1}$ states in $^{97}$Ag.

Let us now come to the electromagnetic observables.
The effective operators needed for the calculation have been derived
as described in Sec. II.
Experimental information on electromagnetic properties in
proton-rich $N=50$ isotones
is very scanty. 
The measured $E2$ transition rates \cite{gorska97,nndc,grzyw98} are compared
with the calculated values in Table IV, where we also report
our predicted $B(E2)$ values for all the states having an
experimental counterpart. As regards the $B(E2;8^+ \rightarrow 6^+)$
in $^{98}$Cd, the two different measured values result from the experiments
of Refs. \cite{gorska97} and \cite{grzyw98}, where this nucleus was produced
by a fusion-evaporation reaction and by fragmentation of a $^{106}$Cd beam,
respectively. While there are some doubts about both these values 
\cite{grzyw98,grawe00}, the one in Ref. \cite{grzyw98}, which corresponds 
to a proton effective 
charge fairly larger than {\it e}, is consistent with that measured for
$^{96}$Pd. From Table IV we 
see that the agreement between experiment and theory
for the $B(E2;8^+ \rightarrow 6^+)$ and
$B(E2;6^+ \rightarrow 4^+)$ in $^{96}$Pd is quite satisfactory,
the calculated values
being only slightly smaller than the observed ones. 
As for $^{98}$Cd, the calculated $B(E2;8^+ \rightarrow 6^+)$ value agrees
with the result of Ref. \cite{grzyw98} within the error bars. 
It is worth noting
that our results do not differ significantly from those obtained
using an effective proton charge $e_{p}^{\rm eff}=1.35e$.

As regards the magnetic observables, only the magnetic moment 
of the $8^+$ state in $^{96}$Pd is known. The measured value is
$10.97 \pm 0.06$ nm \cite{ragha89}, to be compared with the calculated one
10.54 nm.

We have already mentioned in the Introduction that several calculations
have been performed to study the shell-model structure of the $N=50$
isotones.
We only
comment here on the calculation of Ref. \cite{sinatk92} where, assuming
$^{100}$Sn as a closed core,
the two-hole effective interaction was derived by using the
Sussex matrix elements in a
perturbation scheme up to second order without folded-diagram
renormalization.
As pointed out in Sec. II,
the adopted SH energies, as determined from a least-squares fit to the
energies
of the $N=50$, $37 \leq Z \leq 44$ nuclei, are much higher than ours.
In that work, however, attention was focused on nuclei with
$Z=34-46$ and no results were given for $^{98}$Cd and $^{97}$Ag,
for which experimental information has become available only in more
recent times.
We have therefore found it interesting to
perform calculations for these two nuclei using the effective interaction
and the SH energies of Ref. \cite{sinatk92}. We have also calculated a
more complete spectrum of $^{96}$Pd than that given in \cite{sinatk92}.
Hereafter we shall refer to these calculations as Sussex (SUX) calculations.

The $\sigma$ value relative to the SUX calculations for
$^{98}$Cd, $^{97}$Ag, and $^{96}$Pd turns out to be 84, 353, and 218 keV,
respectively.
More explicitly, the experimental position of the positive-parity states
is well reproduced. In particular, the calculated energies of the $2^+$
states in both even isotopes, as well as those of the $\frac{13}{2}^+$
and $10^+$ states in $^{97}$Ag and $^{96}$Pd, come closer to the
experimental
values than those obtained from our calculations. For all other
positive-parity states
the agreement with experiment obtained from SUX and our calculations is
comparable. On the other hand,
the $\frac{17}{2}^-$
and $5^-$ states in $^{97}$Ag and $^{96}$Pd lie 704 and
560 keV above the experimental ones, respectively, and the excitation energies of the first $5^-$ and $\frac{1}{2}^-$
states in $^{98}$Cd and $^{97}$Ag are predicted to be about 3.5 and 1.4 MeV,
which are not consistent with the experimental
information available for the lighter isotones. Furthermore, for $^{97}$Ag
the $p_{3/2}$ and $f_{5/2}$ strengths are concentrated in the
second $\frac{3}{2}^{-}$ and the second $\frac{5}{2}^{-}$ states, which are 
predicted to lie at about 3 and 4 MeV, respectively.
These values are more than 1 MeV higher than ours which come quite close to those obtained by extrapolating the energies of these states from the
lighter isotones. 
>From the above we feel that the values of the SH energies
adopted in Ref. \cite{sinatk92} are overestimated. 
As regards the $p_{1/2}$ level, this conclusion is strongly supported by the fact that, as mentioned above, the calculated energies of the $\frac{17}{2}^-$
and $5^-$ states in $^{97}$Ag and $^{96}$Pd, which are substantially dependent on the position of this level, are largely overestimated. 
On the other hand, we have
verified that decreasing the values of SH energies is not sufficient to
improve the
agreement between theory and experiment on the whole. In fact, while this 
produces a lowering of the negative-parity states it moves the positive-parity
states up to too high an energy. This latter effect, however, might be 
compensated by taking into account folded-diagram renormalization, which
produces in general a compression of the spectra. 
In this context, we may mention that the authors of Ref.
\cite{sinatk92} say 
that the folded-diagram renormalization
would have worked against the outcome of their calculations. Our preceding
remarks indicate that this conclusion could have been turned round had they made a different choice of the SH energies.

\section{Summary and conclusions}
In this paper, we have performed a shell-model study of the $N=50$ isotones
$^{98}$Cd, $^{97}$Ag, and $^{96}$Pd employing a two-hole effective
interaction derived from the Bonn-A nucleon-nucleon potential by means of a
$G$-matrix folded-diagram approach.
We have shown that all the experimental data available for these nuclei
are well reproduced by our calculations. In addition, some of our
predictions, namely the existence of a $5^-$ state in $^{98}$Cd and a ${1
\over 2}^-$ state in $^{97}$Ag  at 2.7 and 0.5 MeV,
respectively, are strongly supported by the experimental information
existing for the lighter isotones.

This work is framed in the context of an extensive program of calculations
aimed at assessing just how accurate a description of nuclear structure
properties can be provided by an effective interaction derived from the $NN$
potential. The quality of the results presented here turns out to be
comparable to that obtained in the $^{132}$Sn and $^{208}$Pb regions where,
however, there is a substantially larger body of experimental data with
which to compare the calculated spectroscopic properties. In particular, we
emphasize that the experimental information existing for the $N=50$ isotones
provides only little guidance to the choice of the SH energies, which
renders it a difficult task. In this situation, our choice has been based
on an analysis of the spectra of $^{98}$Cd, $^{97}$Ag, 
and $^{96}$Pd and on the values of the experimental $\frac{3}{2}^{-}$
and $\frac{5}{2}^{-}$ single-hole energies in $^{89}$Y, $^{91}$Nb,
and $^{93}$Tc. 
We feel, however,  that the unavoidable uncertainty in the adopted
SH energies should not be so severe as to make our findings questionable.

In conclusion, we are confident that the present work may be of stimulus and
help towards  new experimental studies of the proton-rich $N=50$ isotones.

\acknowledgements
\noindent
{This work was supported in part by the Italian Ministero dell'Universit\`a
e della Ricerca Scientifica e Tecnologica (MURST) and by the U.S. Grant
No. DE-FG02-88ER40388. NI thanks the European Social Fund for financial 
support.}

\mediumtext

\begin{figure}
\caption{Experimental and calculated spectrum of $^{98}$Cd}
\end{figure}

\begin{figure}
\caption{Experimental and calculated spectrum of $^{97}$Ag}
\end{figure}

\begin{figure}
\caption{Experimental and calculated spectrum of $^{96}$Pd}
\end{figure}

\mediumtext
\begin{table}
\setdec 0.00
\caption{Calculated energy levels in $^{98}$Cd up to 5 MeV.}

\begin{tabular}{lclc}
$J^{\pi}$ & E(MeV) &$J^{\pi}$& E(MeV)  \\ 
\tableline
$0^{+}$ & 0.0   & $0^{+}$ & 3.499\\  
$2^{+}$ & 1.606 & $3^{-}$ & 3.811\\
$4^{+}$ & 2.117 & $5^{-}$ & 4.381\\
$6^{+}$ & 2.295 & $4^{-}$ & 4.466\\
$8^{+}$ & 2.414 & $2^{+}$ & 4.578\\
$5^{-}$ & 2.705 & $6^{-}$ & 4.594 \\
$4^{-}$ & 3.083 & $7^{-}$ & 4.801\\
\end{tabular}
\label{Table I}
\end{table}

\begin{table}
\setdec 0.00
\caption{Calculated energy levels in $^{97}$Ag up to 3 MeV.}

\begin{tabular}{lclc}
$J^{\pi}$ & E(MeV) &$J^{\pi}$& E(MeV)  \\ 
\tableline
$\frac{9}{2}^{+}$ & 0.0 & $\frac{15}{2}^{+}$ & 2.055  \\
$\frac{1}{2}^{-}$ & 0.528 &$\frac{17}{2}^{+}$ & 2.077    \\
$\frac{7}{2}^{+}$ & 1.007 &$\frac{9}{2}^{-}$ & 2.210     \\
$\frac{5}{2}^{+}$ & 1.342 &$\frac{21}{2}^{+}$ & 2.318    \\
$\frac{13}{2}^{+}$ & 1.501 &$\frac{13}{2}^{-}$ & 2.326   \\
$\frac{11}{2}^{+}$ & 1.593 &$\frac{17}{2}^{-}$ & 2.339    \\
$\frac{3}{2}^{+}$ & 1.648 &$\frac{7}{2}^{-}$ & 2.479    \\
$\frac{9}{2}^{+}$ & 1.766 &$\frac{5}{2}^{-}$ & 2.558    \\
$\frac{5}{2}^{-}$ & 1.801 &$\frac{11}{2}^{-}$ & 2.747    \\
$\frac{3}{2}^{-}$ & 1.814 &$\frac{9}{2}^{+}$ & 2.929   \\
$\frac{3}{2}^{-}$ & 2.024 &$\frac{15}{2}^{-}$ & 2.962    \\
\end{tabular}
\label{Table II}
\end{table}

\begin{table}
\setdec 0.00
\caption{Calculated energy levels in $^{96}$Pd up to 4 MeV.}

\begin{tabular}{lclc}
$J^{\pi}$ & E(MeV) &$J^{\pi}$& E(MeV)  \\ 
\tableline
$0^{+}$ & 0.0   &$4^{-}$ & 3.422\\
$2^{+}$ & 1.678 &$4^{+}$ & 3.559 \\
$4^{+}$ & 2.201 &$3^{-}$ & 3.569 \\
$6^{+}$ & 2.397 &$3^{+}$ & 3.585     \\
$8^{+}$ & 2.507 &$0^{+}$ & 3.619    \\
$5^{-}$ & 2.618 &$3^{-}$ & 3.744   \\
$4^{+}$ & 2.805 &$6^{+}$ & 3.762    \\
$4^{-}$ & 2.991 &$7^{-}$ & 3.827    \\
$2^{+}$ & 3.048 &$8^{+}$ & 3.877    \\
$6^{+}$ & 3.085 &$3^{-}$ & 3.916    \\
$0^{+}$ & 3.161 &$6^{-}$ & 3.920    \\
$5^{+}$ & 3.354 &$10^{+}$ & 3.924    \\
$7^{+}$ & 3.413 & &     \\
\end{tabular}
\label{Table III}
\end{table}

\mediumtext
\begin{table}
\setdec 0.00
\caption{Calculated and experimental $E2$ transition rates
(in W.u.) between yrast states in $^{98}$Cd, $^{97}$Ag, and $^{96}$Pd.}

\begin{tabular}{lccc}
 Nucleus &$J^{\pi}_i \rightarrow J^{\pi}_f$ &
$B(E2)_{\rm calc}$ &$B(E2)_{\rm expt}$  \\
\tableline
 $^{98}$Cd & $2^+ \rightarrow  0^+$    & 3.71  & ~~ \\
 ~~~~~~~~~ & $4^+ \rightarrow  2^+$    & 4.28  & ~~ \\
 ~~~~~~~~~ & $6^+ \rightarrow  4^+$    & 2.98  & ~~ \\
 ~~~~~~~~~ & $8^+ \rightarrow  6^+$    & 1.20  &${1.17^{+0.14}_{-0.11}}^
{\footnotesize{a}}$, ${0.54^{+0.28}_{-0.13}}$ \hspace{-0.08in}
\raisebox{0.04in} {$\scriptstyle{b}$} \\
 $^{97}$Ag & $\frac{13}{2}^+ \rightarrow \frac{9}{2}^+$
& 4.23 & ~~ \\
 ~~~~~~~~~~& $\frac{17}{2}^+ \rightarrow \frac{13}{2}^+$
& 4.15 & ~~ \\
 ~~~~~~~~~~& $\frac{21}{2}^+ \rightarrow \frac{17}{2}^+$
& 2.42 & ~~ \\
 $^{96}$Pd & $2^+ \rightarrow  0^+$    & 6.27  & ~~ \\
 ~~~~~~~~~ & $4^+ \rightarrow  2^+$    & 0.78  & ~~ \\
 ~~~~~~~~~ & $6^+ \rightarrow  4^+$    & 0.51  & $0.78 \pm 0.11$ \\
 ~~~~~~~~~ & $8^+ \rightarrow  6^+$    & 0.20  & $0.34 \pm 0.05$ \\
 ~~~~~~~~~ & $10^+ \rightarrow  8^+$   & 4.29  & ~~ \\
 ~~~~~~~~~ & $12^+ \rightarrow  10^+$  & 3.28  & ~~ \\
\end{tabular}
$^a$ {\footnotesize Reference \cite{grzyw98}.}

$^b$ {\footnotesize Reference \cite{gorska97}.}
\label{table IV}
\end{table}


\begin{references}

\bibitem{andr96} Andreozzi F,  Coraggio L, Covello A, Gargano A,
Kuo T T S, Li Z B and Porrino A 1996 {\it Phys. Rev.} C {\bf 54} 1636

\bibitem{andr97} Andreozzi F, Coraggio L, Covello A, Gargano A,
Kuo T T S and Porrino A 1997 {\it Phys. Rev.} C {\bf 56} R16

\bibitem{cov97} Covello A, Andreozzi F, Coraggio L, Gargano A,
Kuo T T S and Porrino A 1997 {\it Prog. Part. Nucl. Phys.} {\bf 38} 165 

\bibitem{cov98} Covello A, Coraggio L and  Gargano A 1998 {\it Structure
of Nuclei under Extreme Conditions}: {\it Proc.  16th EPS Nuclear
Physics Division Conference (Padova, Italy, 1998)} {\it Nuovo Cimento}  
A {\bf 111} 803 

\bibitem{cor98} Coraggio L, Covello A, Gargano A, Itaco N and 
Kuo T T S 1998 {\it Phys. Rev.} C {\bf 58} 3346

\bibitem{andr99} Andreozzi F, Coraggio L, Covello A, Gargano A,
Kuo T T S and Porrino A 1999 {\it Phys. Rev.} C {\bf 59} 746

\bibitem{cov99} Covello A, Coraggio L, Gargano A, Itaco N and Kuo
T T S 1999 {\it Nuclear Structure 98 (Gatlinburg, Tennessee, 1998)} 
ed C Baktash {\it AIP Conf. Proc.} 481 (New York: AIP) p 56

\bibitem{cor99} Coraggio L, Covello A, Gargano A, Itaco N and Kuo
T T S 1999 {\it Phys. Rev.} C {\bf 60} 064306

\bibitem{machl87} Machleidt R, Holinde K and Elster Ch 1987 {\it Phys. Rep.} 
{\bf 149} 1 

\bibitem{machl89} Machleidt R  1989 {\it Adv. Nucl. Phys.} {\bf 19} 189

\bibitem{jiang92} Jiang M F, Machleidt R, Stout D B and Kuo T T S
1992 {\it Phys. Rev.} C {\bf 46} 910

\bibitem{hjorth95} Hjorth-Jensen M, Kuo T T S and Osnes E 1995 {\it Phys. Rep.}
{\bf 261} 15

\bibitem{gorska97} M. G\'orska {\it et al.} 1997 {\it Phys. Rev. Lett.} 
{\bf 79} 2415

\bibitem{blomq85} Blomqvist J and Rydstr\"om L 1985 {\it Phys. Scr.} {\bf 31} 31

\bibitem{ji88a} Ji X and Wildenthal B H 1988 {\it Phys. Rev.} C {\bf 37}
1256

\bibitem{ji88b} Ji X and Wildenthal B H 1988 Phys. Rev. C {\bf 38} 2849

\bibitem{sinatk92} Sinatkas J, Skouras L D, Strottman D and 
Vergados J D 1992 {\it J. Phys.} G {\bf 18} 1377

\bibitem{ghugre95} Ghugre S S and Datta S K 1995 {\it Phys. Rev.} C {\bf 52} 
1881

\bibitem{johnst97} Johnstone I P and Skouras L D 1997 {\it Phys. Rev.} C 
{\bf 55} 1227

\bibitem{zhang99} Zhang C-H, Wang S-J and Gu J-N 1999 {\it Phys. Rev.}
C {\bf 60} 054316

\bibitem{elliott68} Elliott J P, Jackson A D , Mavromatis H A, 
Sanderson E A and Singh B 1968 {\it Nucl. Phys.} A {\bf 121} 241

\bibitem{krenc76} Krenciglowa E M,  Kung C L, Kuo T T S and Osnes E 1976
{\it Ann. Phys. (N.Y.)} {\bf 101} 154 

\bibitem{kuo96} Kuo T T S 1996 {\it New Perspectives in Nuclear
Structure}:
{\it Proc. 5th Int. Spring Seminar on Nuclear Physics
(Ravello, Italy, 1995)} ed A Covello (Singapore: World Scientific)
p 131

\bibitem{tsai72} Tsai S F  and Kuo T T S 1972  {\it Phys. Lett.} B {\bf 39} 427

\bibitem{shurp83} Shurpin J, Strottman D and Kuo T T S 1983 {\it Nucl. Phys.}
A {\bf 408} 310

\bibitem{kuo81} Kuo T T S, Shurpin J, Tam K C, Osnes E and 
Ellis P J 1981
{\it Ann. Phys. (N.Y.)} {\bf 132} 237

\bibitem{krenc80} Krenciglowa E M  and Kuo T T S 1980  {\it Nucl. Phys.} A
{\bf342} 454

\bibitem{lesu80} Suzuki K and Lee S Y 1980 {\it Prog. Theor. Phys.} 
{\bf 64} 2091

\bibitem{mavro66} Mavromatis H A, Zamick L and Brown G E 1966 {\it Nucl. Phys.}
{\bf 80} 545

\bibitem{krenc75} Krenciglowa E M  and Kuo T T S 1975 {\it Nucl. Phys.} A 
{\bf 240} 195

\bibitem{shamai72} Shamai Y, Ashery D, Yavin A I. Bruge G and 
Chaumeaux A 1972 {\it Nucl. Phys.} A  {\bf 197} 211

\bibitem{brown} Brown B A, Etchegoyen A and Rae W D M 1988 
The computer code OXBASH {\it MSU-NSCL Report} 524

\bibitem{nndc} Data extracted using the NNDC On-Line Data Service from the
ENSDF database, file revised as January 12, 2000, Bhat M R  1992 {\it Evaluated
Nuclear Structure Data File (ENSDF): Nuclear Data for Science and
Technology} ed S M Qaim (Berlin: Springer-Verlag) p 817


\bibitem{grzyw98} Grzywacz R {\it et al.} 1988 {\it ENAM98} ed B M
Sherrill, D J Morrissey and C N Davids {\it AIP Conf. Proc.} 455 
(New York: AIP) p 430

\bibitem{grawe00} Grawe H Private communication.

\bibitem{ragha89} Raghavan P 1989 {\it At. Data Nucl. Data Tables} {\bf 42} 189

\end{references}
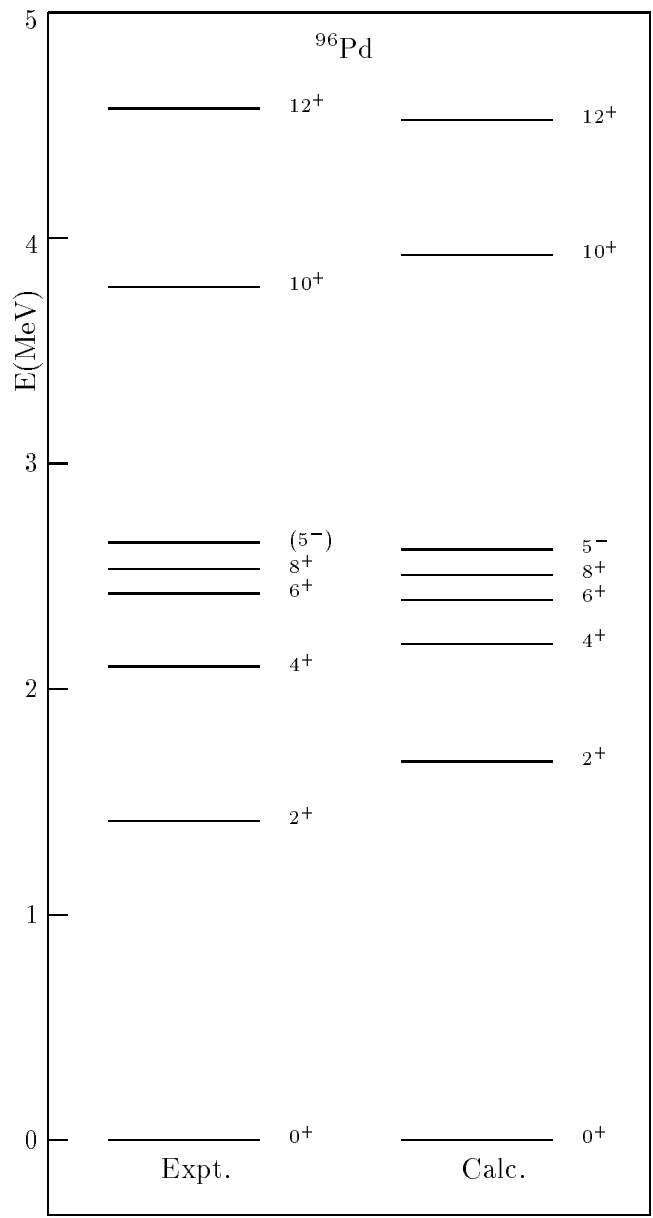
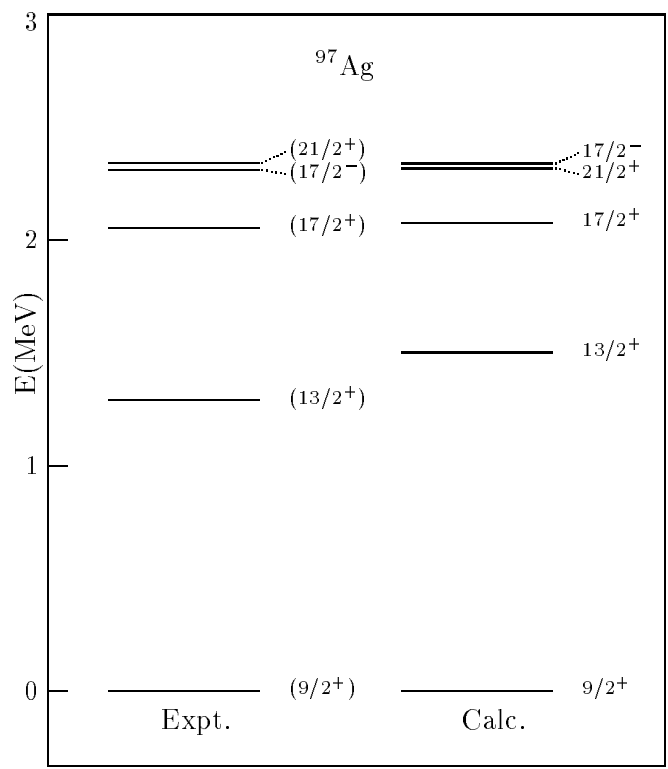
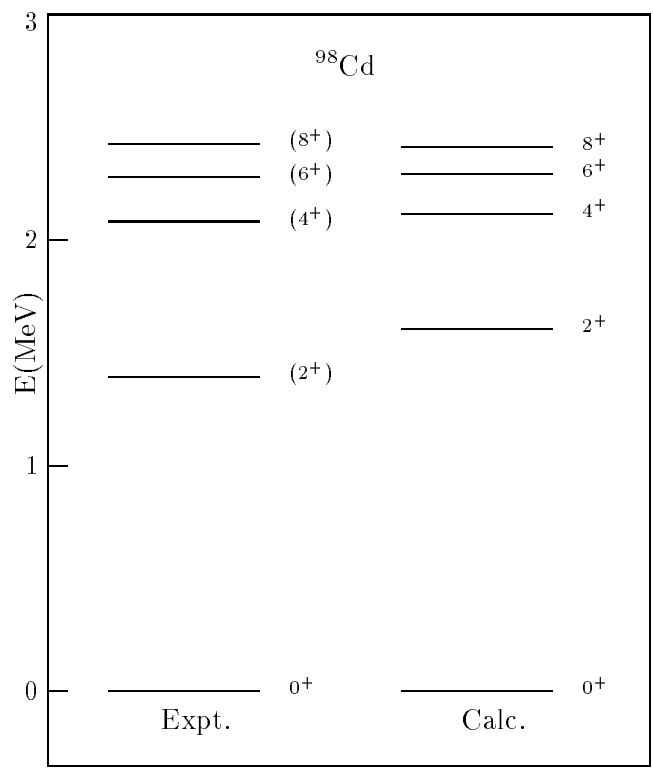
\end{document}